# Observation of Relativistic Electrons Deflection by a Bunch Coulomb Field

**G Naumenko, Yu Popov and M Shevelev**
Tomsk Polytechnic University, Lenina str. 2, Tomsk, 634050, Russia

E-mail: naumenko@tpu.ru

**Abstract.** According to the E.L. Feinberg viewpoint on the possibility of a semi-bare electron state we may expect a state of electron with non-equilibrium Coulomb field during the time range of $\gamma^2 \lambda / c$, where $\gamma$ is the Lorents-factor of electron, $c$ is light velocity and $\lambda$ is the wavelength corresponding to a frequency in Fourier expansion of the Coulomb field of relativistic electron. If this non-equilibrium Coulomb field has a spatial asymmetry, a relativistic electron may be deflected by its own Coulomb field. For the $\gamma^2 \lambda \approx 1$ m we may observe this effect in macroscopic mode when the electrons are grouped in bunches with the population $N_e \approx 10^8$. We present in this paper the experimental observation of the deflection of an electron beam by an asymmetrical non-equilibrium Coulomb field of electron bunch. The experiment was performed on the relativistic electron beam of the microtron of Tomsk Polytechnic University.

## 1. About problem

The shadowing effect – the effect when a part of the Coulomb field of relativistic electron is reflected from a conductive screen, or is absorbed in absorbing screen, and the shadow area appears downstream to the screen, was investigated experimentally in [1]. In [2] was shown that this effect cannot be explained as interference between the electron field and a transition (or diffraction) radiation from the screen, and should be considered as a non-equilibrium state of the electron Coulomb field.

Most clearly, this problem may be represented in the view of the electron field as a field of pseudo-photons. The "pseudo-photon" method proposed by Fermi [3] and developed by Williams [4] is widely used for theoretical studies of electromagnetic processes (see for example [5] and [6]). According to this approach, the field of a charged particle may be replaced by a field of photons, which in this case are called pseudo-photons (in [7] is used the term "virtual quanta"; it should be differed on the same term in the quantum theory). This approach provides a good accuracy for the ultra-relativistic particles when the particle velocity is close to the light velocity (v ~ c) (see [5]) and when the longitudinal electric field component of the particle is negligible. In this case the particle field has the same properties as the field of real photons.

Let us consider the interaction of real photons with a thick conductive mirror of high reflectivity. It is clear that real photons are reflected almost completely, they do not penetrate the target, and they do not excite surface currents on the downstream surface of the target. Because the properties of pseudo-photons are close to the properties of real photons, we may expect that with the passage of electrons close to an edge of a conductive screen (see figure 1) the part of the electron field will be reflected from the screen (so named "backward diffraction radiation" (BDR)) and downstream to the screen this part of electron Coulomb field will be lost ("semi-bare" electron).

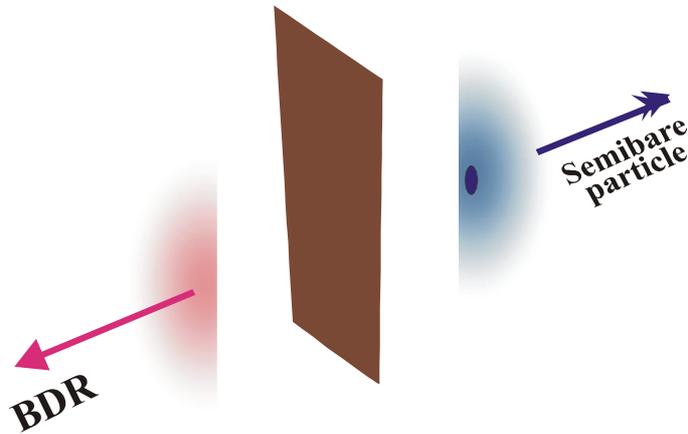

**Figure 1**. The shadowing effect as a reflection of the particle Coulomb field pseudo-photons.

During some time the remaining Coulomb field will have an axial asymmetry, which is the Coulomb field in non-equilibrium state. It is clear that in the further evolution the Coulomb field of the electron will be restored, because far from the screen we always observe electrons with the usual Coulomb field. However during the recovery process the electron may appear in a region of nonzero electric field and this field may deflect the electron like in figure 2.

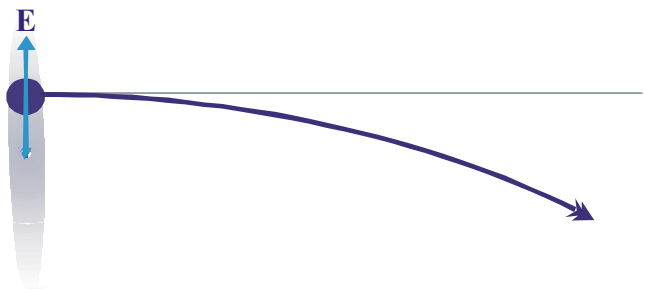

**Figure 2**. Electron deflection by asymmetrical Coulomb field

Of cause, for the single electron this effect is negligible and should be considered in frame of quantum electrodynamics. However, if the electrons are grouped in bunches with the population $10^8$ (like for our accelerator), then for the coherent frequencies this effect may be observable and may be considered in frame of classical electrodynamics.

## 2. Experiment
The experiment was carried out in the extracted electron beam of the Tomsk Nuclear Physics Institute microtron with parameters presented in Table 1.

**Table 1.** Electron beam parameters.

| Electron energy | 6.1 MeV ($\gamma = 12$) | Bunch period | 380 psec |
|---|---|---|---|
| Train duration | $\tau \approx 4$ $\mu$sec | Bunch population | $N_e = 6 \cdot 10^8$ |
| Bunches in a train | $n_b \approx 1.6 \cdot 10^4$ | Bunch length | $\sigma \approx 1.3 \sim 1.6$ mm |

The measurements were performed using the scheme shown in figure 3. Relativistic electrons grouped in bunches of approximately $10^8$ electrons, passing close to the conducting screen.

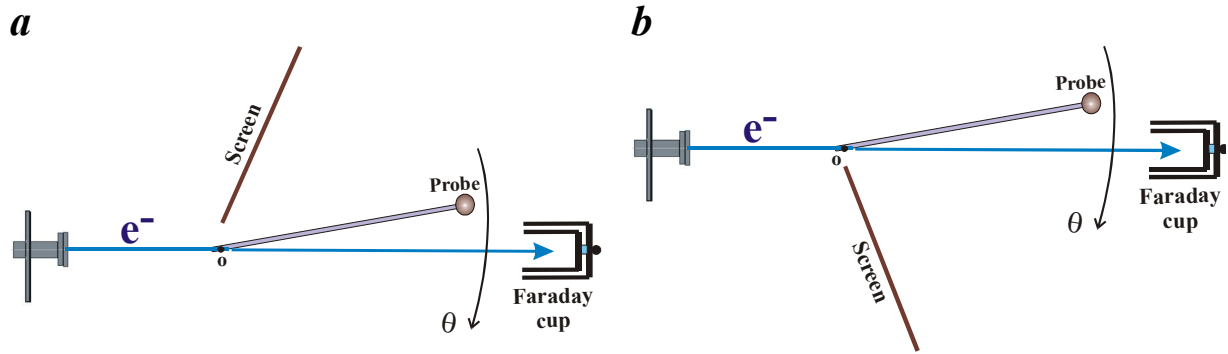

**Figure 3**. The cheme of the angular beam profile measurement. *a* –for left screen position and *b* – for right screen position.

The screen was inclined at the angle $30^0$ to exclude the influence of backward diffraction radiation on the electron beam. The minimal distance $h_0$ between the screen and electron beam was chosen using Faraday cup response dependence on $h$ (see figure 4). The screen position $h_s$ was chosen so that the electron losses on the screen are to be less than 5%.

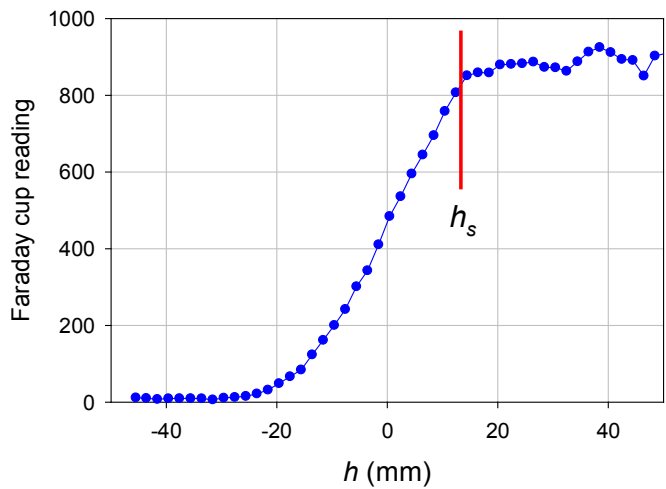

**Figure 4**. Dependence on impact parameter, and choosing of the working screen position.

The electron beam angular profile was measured using the probe - the screened metallic stick which absorbs a part of electron of the beam. The probe may be rotated around the point o (see figure 3) to measure the beam profile. The sample of the measured beam profile is shown in figure 5.
For the calculation of the beam position we had approximated the upper third part of the angular dependence (see figure 5) by the Gaussian dependence with calculation of the position error with the confidence probability equal to 95%.

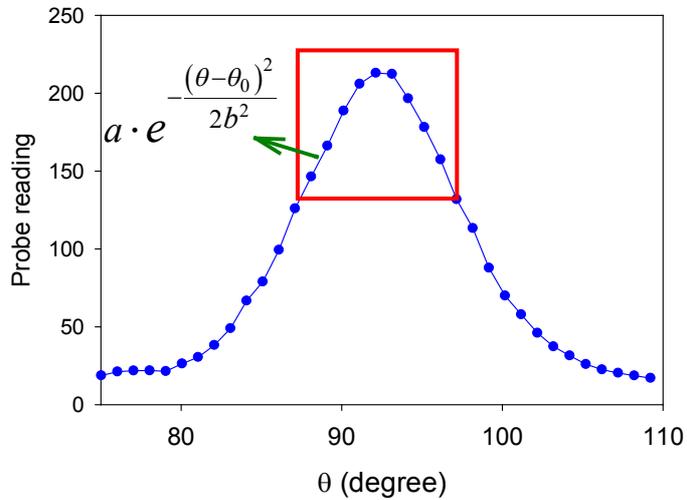

**Figure 5**. The sample of the angular profile of the electron beam. The upper third part of dependence (selected by rectangle) is approximated by gaussian function and the fitting of this function position $\theta_0$ is used as estimation of the beam position.

The effect is so small, that the apparatuses drift should be taken into account. To exclude the apparatuses drift, the measurements were performed by the sequential repetition of the tetrad of measurements:
1) Profile measurement without the screen,
2) Measurement with the screen in the left position (figure 3*a*),
3) Measurement without the screen,
4) Measurement with the screen in the right position (figure 3*b*),

The results of the beam position sequential measurements are shown in figure 6.

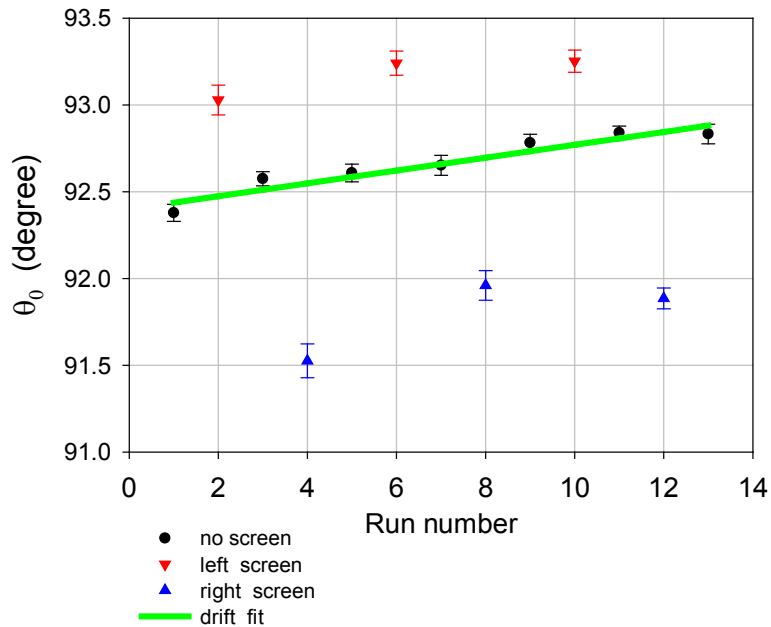

**Figure 6**. Series of the sequential measurements of the beam deflection. The solid line is the linear fit of the drift.

After the linear fitting, we had subtracted the drift from measured data and presented the final results in figure 7.

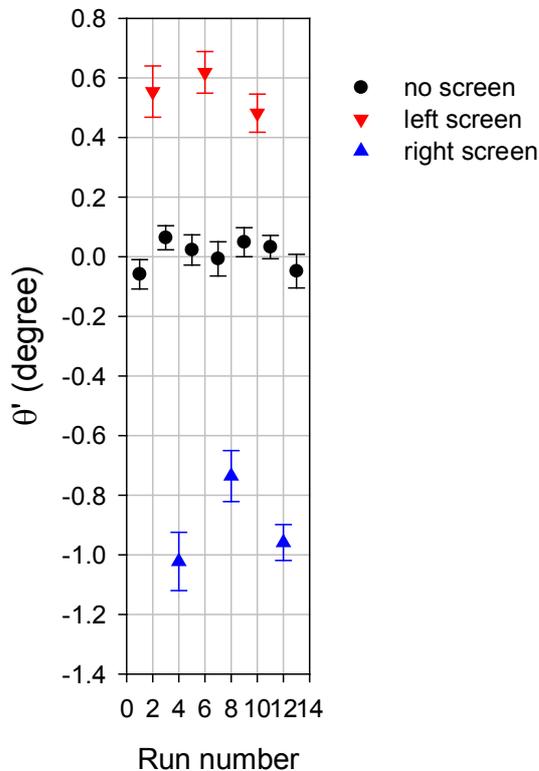

**Figure 7**. Beam deflection with accounting of the drift.

We can see in figure 7, that the effect of electron beam deflection far exceeds the experimental error. For the checking of a possible influence of the backward coherent diffraction radiation from the conductive screen on the electron beam, we had replaced the conductive screen by the absorbing one. In limits of the experimental error the same results were obtained. So, the observed beam deflection cannot be explained as the influence of the backward coherent diffraction radiation and may be explained only as a bunch deflection by its own Coulomb field. This effect is close to generation of electron beams carrying orbital angular momentum [8], but in our case transversal momentum of a beam is generated.

There is a question, what is difference between the deflection effect and the well-known recoil effect? On our opinion both these effects are the different viewpoints on the same phenomenon, and no principal difference is between these viewpoints. The recoil effect viewpoint is based on the energy and momentum conservation law. Therefore results may be obtained only in terms of a finite state (in classical notion – in far zone), and a dynamics may not be analyzed in contrast to the deflection effect viewpoint. Also there is a question, how in case of absorbing screen the recoil effect may be considered? The nature of this phenomenon is more understood in frame of deflection effect viewpoint.

Finally we can conclude, that the bunch deflection at the angle $\approx 0.7 \pm 0.06^O$ by its own Coulomb field was registered for the relativistic electron with energy 6.2 MeV on the Tomsk microtron.

**Acknowledgment**
This work was partly supported by the warrant-order 1.226.08 of the Ministry of Education and Science of the Russian Federation.